\documentclass{article}
\usepackage{dirtytalk}
\usepackage{spconf,amsmath,graphicx,xcolor}
\usepackage[british]{babel}
\usepackage{multirow}
\usepackage{epstopdf}
\ninept

\title{Simulating dysarthric speech for training data augmentation in clinical speech applications}
\name{Yishan Jiao$^1$, Ming Tu$^1$, Visar Berisha$^{1,2}$ and Julie Liss$^1$}
\address{$^1$Department of Speech and Hearing Science\\
$^2$ School of Electrical, Computer, and Energy Engineering\\
Arizona State University}
\begin{document}
%
\maketitle
\begin{abstract}
Training machine learning algorithms for speech applications requires large, labeled training data sets. This is problematic for clinical applications where obtaining such data is prohibitively expensive because of privacy concerns or lack of access. As a result, clinical speech applications are typically developed using small data sets with only tens of speakers. In this paper, we propose a method for simulating training data for clinical applications by transforming healthy speech to dysarthric speech using adversarial training. We evaluate the efficacy of our approach using both objective and subjective criteria. We present the transformed samples to five experienced speech-language pathologists (SLPs) and ask them to identify the samples as healthy or dysarthric. The results reveal that the SLPs identify the transformed speech as dysarthric 65\% of the time. In a pilot classification experiment, we show that by using the simulated speech samples to balance an existing dataset, the classification accuracy improves by about 10\% after data augmentation.
\end{abstract}
\section{Introduction}
\label{sec:intro}

Recent studies in machine learning have shown that models built from large data sets can achieve extraordinary performance by mining data-driven features directly from the data. Take large vocabulary continuous speech recognition as an example: databases consisting of thousands of hours of speech from many individuals that cover the large variability in speaking style, environment, speaker age, etc., are required to train powerful deep neural networks (DNNs)-based acoustic models \cite{yu2017recent}.

For consumer applications, speech samples can be collected efficiently on a large scale; however for clinical applications of speech analytics, healthy speech samples have only limited utility. For example, if our aim is to build speech-based assistive technology for patients with amyotrophic lateral sclerosis (ALS), simple application of models trained on healthy speech fail even under moderate dysarthria \cite{tu2016relationship}. (Speech produced by people with ALS ranges from mildly slurred and hypernasal, to severely unintelligible as the disease progresses.) Other clinical speech applications that require large labeled training sets include automatic detection of speech disorders \cite{fonseca2007wavelet}\cite{schuller2012interspeech}, intelligibility assessment \cite{kim2015automatic}\cite{middag2009automated}, automatic recognition of disordered speech \cite{rosen2000automatic}\cite{green2003automatic}, automated acoustic measures of speech disorders \cite{jiao2015convex}\cite{jiao2017interpretable}, etc.

Unlike healthy speech, the collection of pathological speech takes longer to conduct and can be more sensitive to other factors, such as variable recording conditions, uncontrolled body movements, unbalanced samples across speakers and diseases, etc. In the literature, there only exist a few publicly-available datasets that are relatively large (e.g., the Nemours database \cite{menendez1996nemours} and the TORGO database \cite{rudzicz2012torgo}); but most researchers opt to collect their own small-scale datasets tailored to their needs. Due to a lack of data, machine learning models used in the study of pathological speech are typically limited to simple unsupervised metrics \cite{berisha2017float}, or flat supervised models \cite{rudzicz2011articulatory}\cite{tu2017objective}. When deep learning models are used \cite{tu2017interpretable}, their solution space is typically constrained using other criteria for better generalization.

Our aim in this paper is to generate simulated dysarthric speech via a model that transforms healthy speech to dysarthric speech so as to augment existing datasets for training large scale machine learning models. We restrict our analysis to speech from people with ALS, a rapidly progressing neurodegenerative disease. Machine learning models based on speech are particularly useful for this group of patients for building new assistive devices that generalize well across disease conditions and patient speaking styles.

We propose to use voice conversion methods to transform healthy speech to ALS dysarthric speech. The method includes speaking rate modification using PSOLA, spectral feature transformation using adversarial training, and pitch modification using a linear transformation. We conducted objective and subjective evaluation to examine whether the simulated ALS speech matches true ALS speech in both the acoustic and perceptual domain. Furthermore, we demonstrate the utility of data augmentation on a classification task. In the remaining parts of this paper, Section \ref{sec:method} introduces the proposed transformation framework and experimental settings. The objective and subjective validation of the method are presented in Section \ref{sec:result} along with a pilot experiment of data augmentation using the simulated samples. We conclude with a discussion of our future plans in Section \ref{sec:discussion}.

\textbf{\emph{Relation to previous work:}} Voice conversion is a technique to modify a source speaker's speech to be perceived as if a target speaker had spoken it \cite{kain1998spectral}. This paper is motivated by this idea, however the source and target speakers in our study are a group of healthy speakers and a group of ALS speakers. Speech transformation of dysarthric speech has also been previously studied \cite{kain2007improving, rudzicz2011acoustic}. In contrast to these studies, the present study attempts to transform healthy speech to pathological speech for the purpose of data augmentation during model training rather than improving intelligibility. Adversarial training has been recently explored in the field of voice conversion \cite{hsu2017voice}\cite{kaneko2017generative_1}\cite{kaneko2017generative_2}. In the most recent work in \cite{hsu2017voice}, the authors used variational autoencoding Wasserstein generative adversarial network (VAW-GAN) to transform speech features, with the assumption that there is a latent variable representing the common phonetic content. However, this assumption fails for ALS speech, where there is a significant deviation from healthy in the acoustic representation of phonemes. As an alternative, we use deep convolutional generative adversarial networks (DCGANs) that has been used for speech synthesis postfiltering \cite{kaneko2017generative_1}\cite{kaneko2017generative_2} to transform speech features.

\begin{figure}[t]
    \centering
    \includegraphics[width=3.2in]{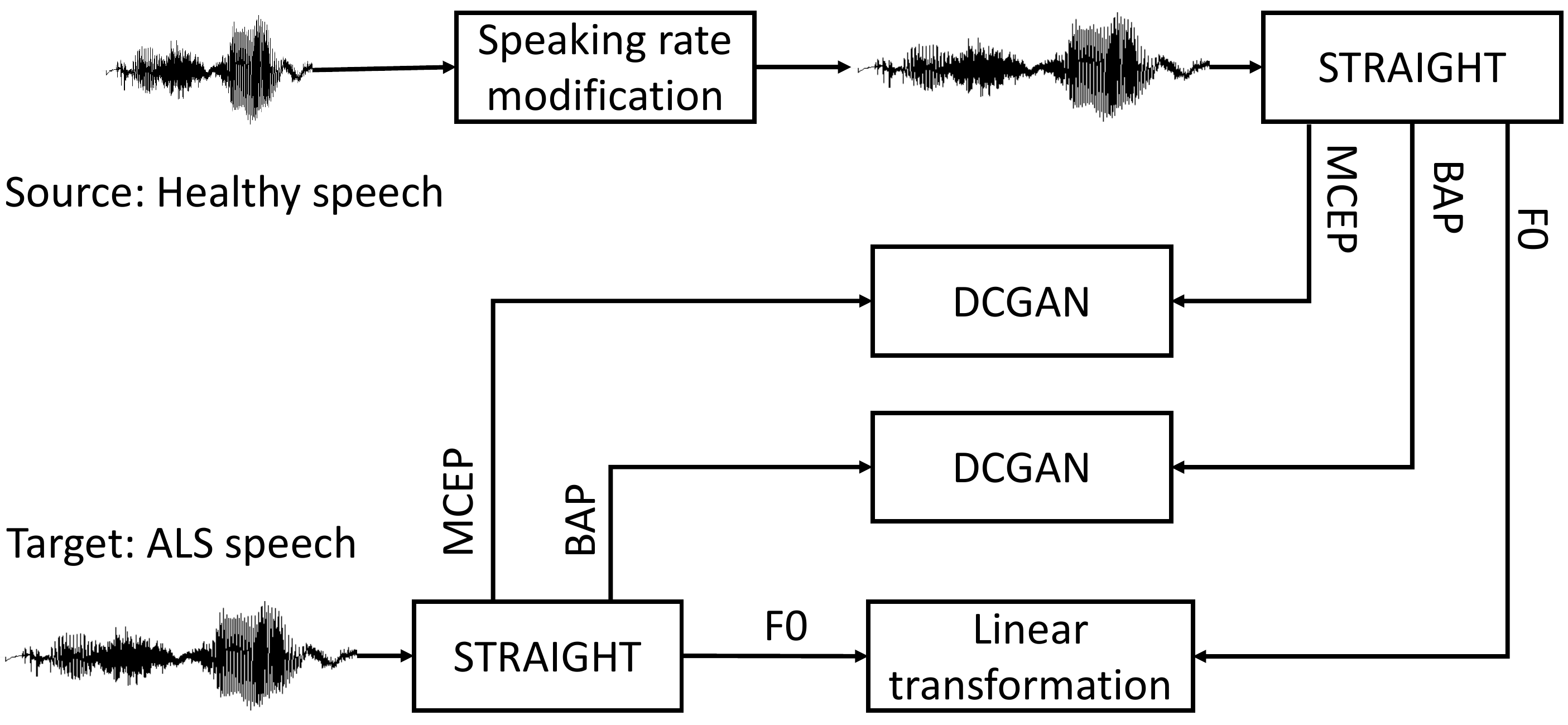}
  \caption{Proposed transformation frame.}
  \centering
  \label{fig:framework}
\end{figure}

\section{Proposed method}
\label{sec:method}
We begin with the assumption that we have a group of ALS speakers with similar perceptual symptoms (e.g. reduced articulatory precision) we aim to model and a group of healthy speakers with a distinct speaking style (e.g. a regional accent). Our aim is to superimpose the ALS symptoms to the speakers with different speaking styles in an attempt to model the variation induced by the disease \emph{and} the variation in speaking style. This allows us to artificially expand the training set in machine learning applications.

The proposed transformation framework is shown in Figure \ref{fig:framework}. Different from traditional voice conversion, the source `speaker' in our study is a group of healthy speakers, and the target `speaker' is a group of ALS speakers as described above. We assume that speakers from both groups read the same materials and we build a mapping between \emph{each} healthy / ALS speaker pair. Suppose there are $N$ healthy speakers and $M$ ALS speakers. Each speaker reads the same $P$ sentences. The paired samples for training are denoted by $\left\{S_{i,k}, T_{j,k}\right\}$, where $S_{i,k}$ and $T_{j,k}$ are the source and target speaker, respectively, with $i=1,2,...,N$, $j=1,2,...,M$, and $k=1,2,...,P$. By performing the training over groups of speakers, we expect that the model will superimpose the ALS symptoms to healthy speech rather than learning a specific speaker's pattern.

People with ALS suffer from mixed flaccid spastic dysarthria, characterized by slow speech rate, imprecise phoneme articulation, hypernasality, monopitch, breathiness, and a harsh voice  \cite{duffy2013motor}. To capture these characteristics, we propose the three-step conversion strategy outlined in Figure \ref{fig:framework}: 1) speaking rate modification; 2) spectral feature transformation using DCGANs; 3) pitch modification. We describe the details of these below.

\subsection{Speaking rate modification}
  The first transformation is to modify the speaking rate of the healthy speech to match the rate of ALS speech. During training, we match each healthy speech sample to the length of its paired ALS speech sample. During testing, each healthy speech sample is modified to a reference value. In our study, the rate of ALS speech was twice as slow on average as the rate of the healthy speech. As a result, for test speech samples, we stretch them to double their lengths. The change in speaking rate is performed by using the open-source Praat Vocal Toolkit \cite{toolkit}, which uses PSOLA to change the duration of the speech while preserving the pitch.

\begin{figure}[t]
    \centering
    \includegraphics[width=3.2in]{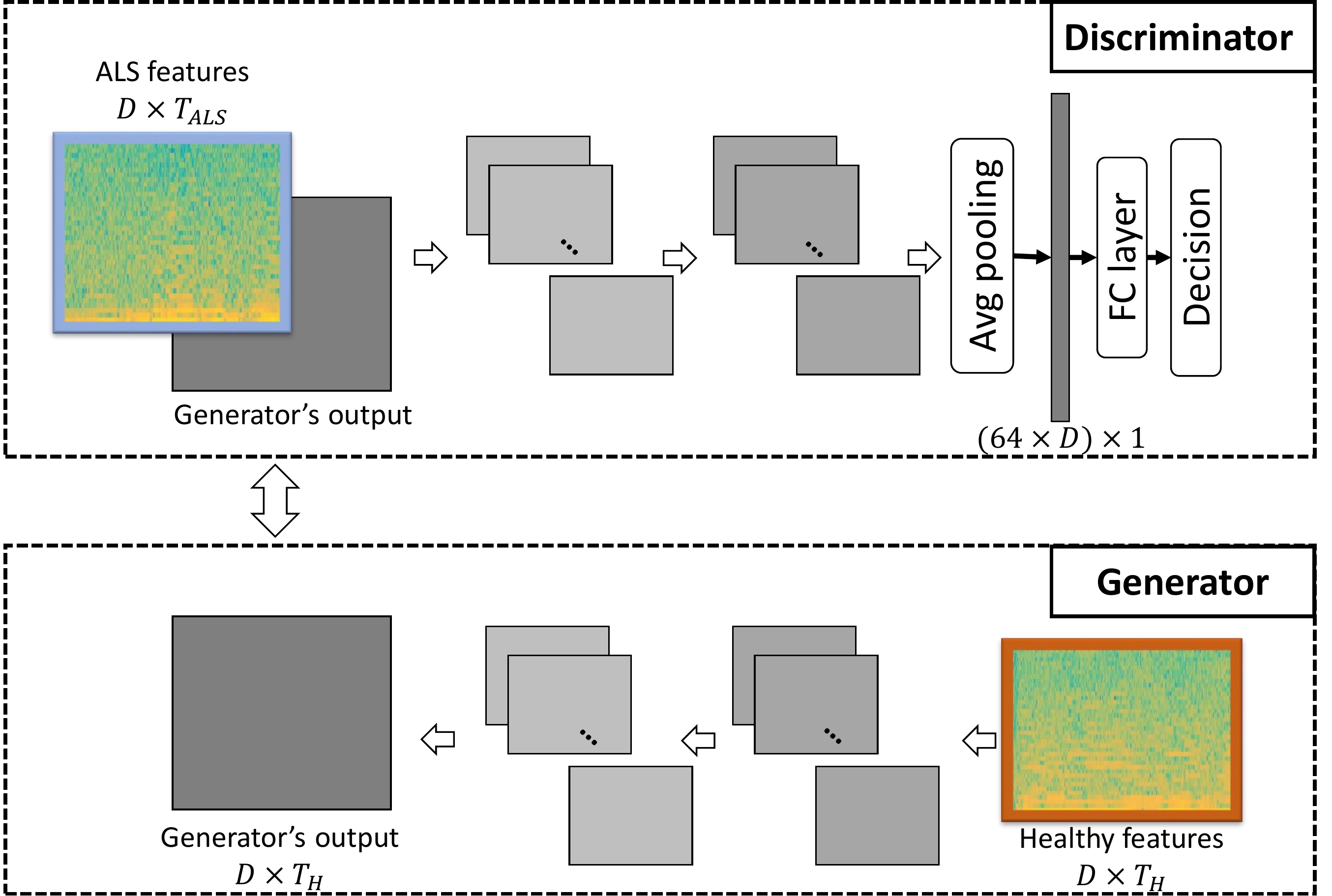}
  \caption{The structure of DCGAN model used in the presented study.}
  \centering
  \label{fig:dcgan}
\end{figure}

\subsection{Speech feature extraction}
 As Figure \ref{fig:framework} shows, after rate modification, we transform the spectral and pitch features extracted by the STRAIGHT analysis \cite{kawahara1999restructuring}. The fundamental frequency $F_0$, spectrogram (SP), and the aperiodic spectrum (AP) are extracted from the ALS speech samples and the rate-modified healthy samples. The SP and AP features are then transformed to 39-dimensional mel-cepstral coefficients (MCEPs) and 24-dimensional band-aperiodicity parameters (BAPs), respectively \cite{sptk2009speech}. We expect that the ALS speech characteristics of imprecise phoneme articulation, hypernasality, and breathiness can be modeled by transforming the MCEPs; harsh vocal quality can be modeled by transforming aperiodic components;  and monopitch can be modeled by modifying $F_0$.

\subsection{Spectral feature transformation}
The spectral features (MCEPs and BAPs) are transformed using adversarial training. Generative adversarial networks (GAN) \cite{goodfellow2014generative} are a machine learning strategy which uses a combination of discriminative and generative models. The generative model tries to generate samples similar to the target, while the discriminative model tries to distinguish between the distribution of generated samples and actual samples. By training the two models simultaneously, it is expected that the generated samples become indistinguishable from the target sample.

In our study, we take advantage of adversarial training and use DCGANs \cite{radford2015unsupervised} to transform healthy speech to ALS speech. The structure of the model we used is shown in Figure \ref{fig:dcgan}. The lower panel is the generator while the upper panel is the discriminator. The architectures of the multilayer convolutional neural network (CNN) in the generator and the discriminator are shown in Table \ref{table:gan}. Within each batch at both the discriminator and the generator, we zero-pad the speech samples until all are of identical temporal length. The input dimension $D$ is the dimension of the spectral features and $T_\mathrm{H}$ and $T_{\mathrm{ALS}}$ are the sequence length (with zero padding) for the healthy and ALS speakers, respectively. In both the generator-CNN (G-CNN) and the discriminator-CNN (D-CNN), convolution with padding is used to keep the input and output dimensions the same. For the generator, the input is the healthy speech feature sequence extracted from an utterance (healthy features shown with a red border in the figure), followed by a multilayer CNN (G-CNN). The output of the generator, which has the same size as the generator input, is sent to the discriminator along with the content-parallelled ALS speech feature sequence (ALS features shown with a blue border in the figure). For the discriminator, a multilayer CNN (D-CNN) is connected to the input feature sequence. The D-CNN processes the ALS and transformed speech in the same fashion. Average pooling is used in the D-CNN to process batches with different lengths by temporally averaging the output of the last convolutional layer. Then flattening is applied to concatenate the outputs of average pooling, resulting in a $64D$-dimensional vector. A fully connected layer is used to make the binary classification decision at the discriminator.

Activation functions for the convolution layer are rectified linear unit (ReLU), and the activation function for the fully-connected layer is a sigmoid. The model was trained using Tensorflow \cite{abadi2016tensorflow}. After hyperparameter tuning, the batch size was set to 32, the learning rate was set to 0.00006 with the Adam optimizer \cite{kingma2014adam}, and the training epoch was 25.

\begin{table}[t]
\centering
\bigskip\caption{The structure of DCGAN.}
\resizebox{\columnwidth}{!}{
\label{table:gan}
\begin{tabular}{ccc} \hline
& Generator-CNN & Discriminator-CNN \\ \hline
Input & $D \times T_{H}$ & $D \times T_{ALS}$(or $T_{H}$) \\ \hline
Conv1 & \begin{tabular}[c]{@{}c@{}}8 conv with 5 $\times$ 5 kernel\\ size and 1 $\times$ 1 stride\end{tabular} & \begin{tabular}[c]{@{}c@{}}8 conv with 5 $\times$ 5 kernel\\ size and 2 $\times$ 2 stride\end{tabular} \\ \hline
Conv2 & \begin{tabular}[c]{@{}c@{}}8 conv with 5 $\times$ 5 kernel\\ size and 1 $\times$ 1 stride\end{tabular} & \begin{tabular}[c]{@{}c@{}}16 conv with 5 $\times$ 5 kernel\\ size and 2 $\times$ 2 stride\end{tabular} \\ \hline
Conv3 & \begin{tabular}[c]{@{}c@{}}1 conv with 5 $\times$ 5 kernel\\ size and 1 $\times$ 1 stride\end{tabular} & \begin{tabular}[c]{@{}c@{}}32 conv with 5 $\times$ 5 kernel\\ size and 2 $\times$ 2 stride\end{tabular} \\ \hline
Conv4 & N/A & \begin{tabular}[c]{@{}c@{}}64 conv with 5 $\times$ 5 kernel\\ size and 2 $\times$ 2 stride\end{tabular} \\ \hline
Output & $D \times T_{H}$ & $64 \times D \times T_{ALS}$(or $T_{H}$) \\ \hline
\end{tabular}}
\end{table}

\subsection{Pitch modification}
Pitch is modified using a linear transformation. An important characteristic of ALS speech is reduced pitch variation. We model this reduction in variation through a linear transformation,
\begin{equation}
\label{eq:f0}
{F_0}^{\mathrm{trans}}(i) = (F_0(i) - \bar{F_0})*\alpha + \bar{F_0}
\end{equation}
where $F_0(i)$ is the estimated nonzero pitch of the healthy speaker for frame $i$, and $\alpha = {\bar{\sigma}_{{F_0}_{ALS}}} /\ \sigma_{F_0}$ is the ratio of standard deviations between the average of ALS speakers and the current healthy speaker.

\subsection{Experimental settings}
We use a subset of a dysarthric speech dataset collected in the Motor Speech Disorders Laboratory at Arizona State University. It consists of speech samples collected from 8 ALS speakers and 8 healthy speakers (4 females and 4 males for each). The dysarthric severity of the ALS speakers ranges from moderate to severe.  Each speaker read the same 80 short phrases (6 syllables in each) in English \cite{liss2009quantifying}. We used the first 70 phrases for training and the other 10 for testing. The training data were organized in a speaker independent style within gender, which means that each female/male healthy speaker was mapped to each of the female/male ALS speakers. Therefore, the total number of training samples including both females and males was 2240, and the number of test samples was 80.

During the training stage, we modify the speaking rate of the healthy samples to match their paired ALS speech. Two DCGAN models were trained on the MCEP and BAP features, respectively. The standard deviation of the pitch contour for each ALS speech sample was calculated. During the test stage, the duration of each healthy speech sample was modified to its double length. The MCEPs and BAPs were transformed using the trained DCGAN models, and the $F_0$ was modified based on Equation \ref{eq:f0}. STRAIGHT synthesis was used to reconstruct the speech signal using the modified $F_0$ and the transformed spectral features.

\section{Results}
\label{sec:result}
\subsection{Objective evaluation}
Objective evaluation was performed to examine whether the generated speech is acoustically similar to true ALS speech. First, we compared the distance between the untransformed and the transformed speech features (MCEP, BAP, and F0) of the test speech samples to true ALS speech (ALS speech samples in training data) using the nonparametric $D_p$ divergence measure \cite{berisha2016empirically}. In order to obtain an unbiased measure, we used bootstrap sampling to ensure the number of samples selected from each of the estimated groups (ALS, healthy, transformed) are the same. The sampling process was done 50 times, and the final measure was obtained by averaging over all trials. The results are shown in Table \ref{tab:res} (\textit{$D_p$-divergence}). A smaller value of the $D_p$ divergence implies that data from the two classes are more similar. The results show that the transformed speech is more similar to true ALS speech than the healthy speech ($p<0.01$).

Second, a support vector machine (SVM) was built to distinguish true ALS speech and healthy speech (using samples in the training set) based on the features we developed previously for representing the characteristics of different types of dysarthric speech \cite{wisler2014domain}. The features include: 1) long-term energy spectrum (LTAS), which captures atypical average spectral information in the signal, related to nasality, breathiness, and aypitcal loudness variations of speech; 2) statistics of Mel-frequency ceptral coefficients (MFCC), which are related to articulation; 3) correlation structure features that capture the evolution of vocal tract shape and dynamics at different time scale via auto- and cross- correlation analysis of formant tracks and MFCC. Since speaking rate is an obvious characteristic to separate ALS speech from healthy speech, we excluded features related to rhythm when building the classifier. After training, the classifier was applied to the untransformed test speech samples (healthy) and their transformations. The error rate of the resulting classifier is shown in Table \ref{tab:res} (\textit{SVM Classification Results}). We can see that the model trained on true ALS and healthy speech classifies the transformed samples as ALS 37.5\% of the time. In contrast, the untransformed healthy speech samples are only classified as ALS 2.1\% of the time.

\vspace{-0.25cm}
\subsection{Subjective evaluation}
Subjective evaluation was performed to examine whether the generated speech was perceptually similar to ALS speech. Five certified SLPs with 12 years clinical experience on average, who routinely work with dysarthric patients, were invited to make a judgement on 20 of the transformed speech samples. The 20 samples were randomly selected from the 80 out-of-sample transformed phrases. The provided instructions were: ``Please determine if the speech sample sounds more like ALS or Healthy speech. When you make your choice, please consider if the speaker shows symptoms of ALS.''

Our transformation procedure induces audible artifacts in the speech. To determine whether these artifacts impact an SLP's decision, we mixed 10 control samples (5 healthy, 5 ALS) with the 20 transformed speech samples. The control samples were generated in the following way: we used the proposed voice conversion framework to transform each control ALS speaker to another ALS speaker, and each control healthy speaker to another healthy speaker. This procedure induces similar artifacts on the control speech samples. The order of the samples were randomized for each listener.

For each speech sample, we collected 5 labels from the 5 SLPs and the results were calculated based on all labels on all samples. The results are shown in the subjective evaluation session of Table \ref{tab:res}. The first row shows the accuracy of the SLPs perceptual classification on the 10 control healthy and ALS samples (with artifacts induced by the transformation). There was strong consensus on these 10 samples among clinicians, with samples correctly classified 98\% of the time. This indicates that the perceptual artifacts induced by the transformation did not impact the ability of the clinicians to correctly classify the speech samples.

The second row shows the percentage of the time that the transformed speech samples were perceived as ALS. However, we noticed that there was consensus (at least 4 same labels) on some of the speech samples, but not on the others. Clinicians also said that it was difficult to make a judgement sometimes. We assumed that for those without consensus, clinicians may make a random guess between ALS and healthy. Therefore, we removed those samples and calculated the result again only when there was consensus (shown in the third row). Across all samples, 65\% of the time clinicians classified the transformed speech samples as ALS; for instances where there was consensus among the clinicians, they were classified as ALS 76\% of the time. The feedback provided by one of the participating clinicians is as follows:

\say{\textit{It was a difficult choice for some of them! Some samples sounded disordered, but not necessarily like ALS patients. I defaulted to ALS if rate was slow and I could not get the phonetic content. Some of the features I detected I would describe as slowed rate, altered resonance (hypernasality), breathy vocal quality, and articulatory imprecision. There were some that had an unnatural/robotic/tinny quality that were deviant but not ALS-like.}}

This quote highlights both the benefits and the drawbacks of the proposed method. It is clear that some of the samples exhibit appropriate reductions in articulation precision, unusual resonance, and breathy characteristics; however the transformation method also produces audible artifacts in the speech signal that must be addressed.

\begin{table}[t]
\centering
\caption{Objective and subjective evaluation results}
\label{tab:res}
\begin{tabular}{|l|l|c|l|}
\hline
\multirow{6}{*}{\rotatebox[origin=c]{90}{\textbf{\begin{tabular}[c]{@{}c@{}}Objective \\ evaluation\end{tabular}}}} & \multicolumn{3}{c|}{\textbf{\textit{$D_p$-divergence}}} \\ \cline{2-4}
 & Healthy vs. ALS & \multicolumn{2}{c|}{0.669} \\ \cline{2-4}
 & Transformed vs. ALS & \multicolumn{2}{c|}{0.552} \\ \cline{2-4}
 & \multicolumn{3}{c|}{\textbf{\textit{SVM Classification Results}}} \\ \cline{2-4}
 & Healthy classified as ALS & \multicolumn{2}{c|}{2.1\%} \\ \cline{2-4}
 & Transformed classified as ALS & \multicolumn{2}{c|}{37.5\%} \\ \hline \hline
\multirow{3}{*}{\rotatebox[origin=c]{90}{\textbf{\begin{tabular}[c]{@{}c@{}}Subjective\\ evaluation \ \ \ \ \end{tabular}}}} & \begin{tabular}[c]{@{}l@{}}Accuracy on control healthy \\ and ALS samples\end{tabular} & \multicolumn{2}{c|}{98\%} \\ \cline{2-4}
 & \begin{tabular}[c]{@{}l@{}}Percentage of transformed \\ perceived as ALS\end{tabular} & \multicolumn{2}{c|}{65\%} \\ \cline{2-4}
 & \begin{tabular}[c]{@{}l@{}}Percentage of transformed \\ perceived as ALS with consensus\end{tabular} & \multicolumn{2}{c|}{76\%} \\ \hline
\end{tabular}
\end{table}

\vspace{-0.25cm}
\subsection{Pilot experiment using data augmentation}
To test if the proposed data augmentation method is effective in improving the performance of machine learning models, we designed a pilot experiment to distinguish between ALS speech and ataxic speech. Ataxia is a neurological disorder resulting from damage to the cerebellar control circuit. Ataxic speech has some common characteristics with ALS speech, such as imprecise phoneme articulation, slowed speaking rate, monopitch, and a harsh voice. It also has its own distinct characteristics, including equal and excess stress, irregular articulatory breakdown and excessive loudness variations.

In our dataset, there are 16 ataxic speakers and 8 ALS speakers with each speaker reading the same 80 phrases. Studies have shown that machine learning classification algorithms are sensitive to unbalanced data. Therefore, we balanced the existing dataset by adding more ALS samples which were transformed from healthy speech using the proposed conversion method. The samples were transformed from the 8 healthy speakers in our dataset.

\begin{figure}[t]
    \centering
    \includegraphics[width=3in]{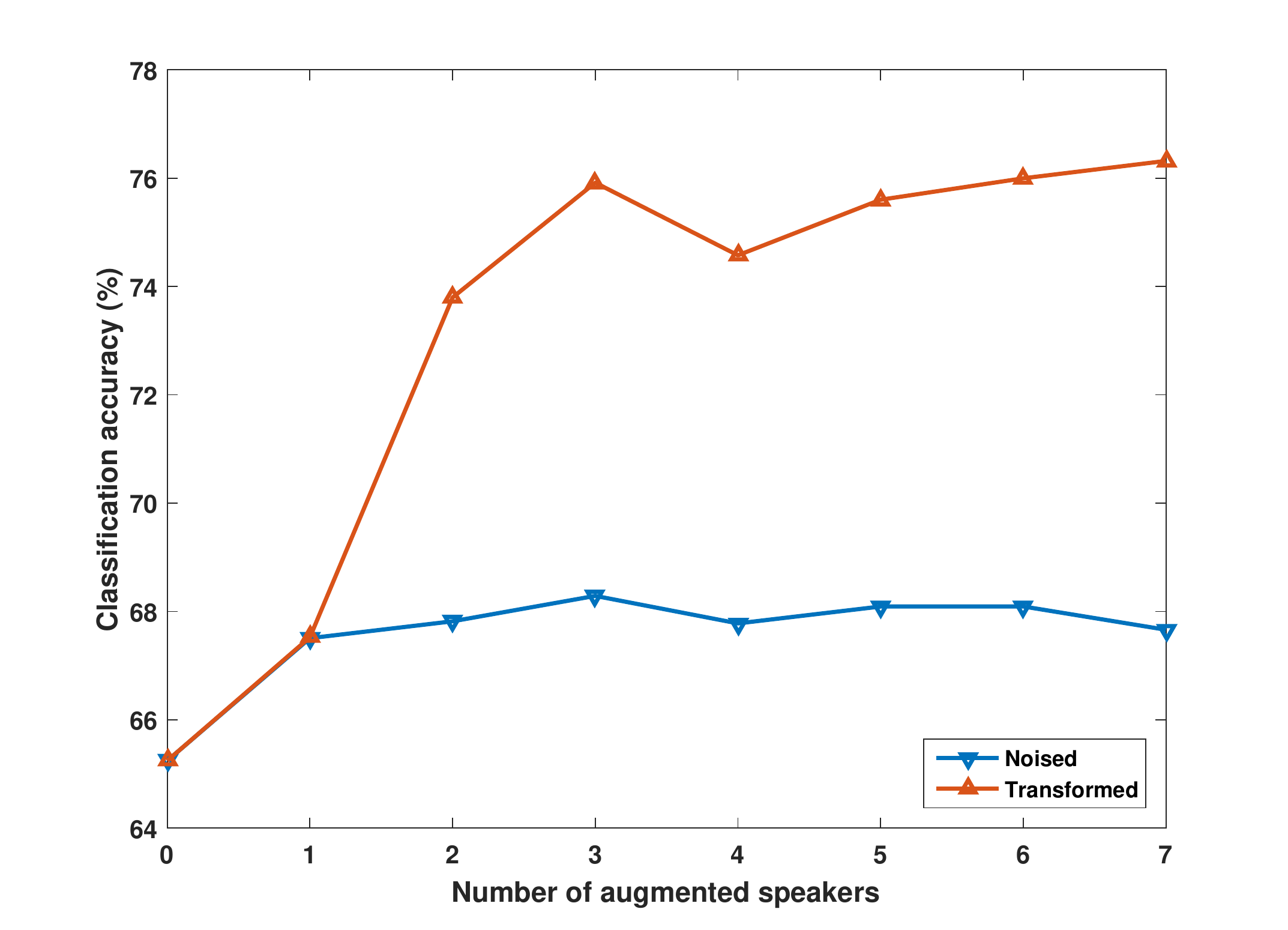}
  \caption{Classification accuracy after data augmentation.}
  \centering
  \label{fig:class_acc}
\end{figure}

An SVM model was built based on the same set of features used in the objective evaluation, plus a rhythm feature, the envelope modulation spectrum (EMS), which is a useful indicator of atypical rhythm patterns in pathological speech. The model was trained by iteratively adding one additional simulated speaker to the unbalanced data set. Leave one-speaker (from the original dataset) out cross-validation was used to evaluate the performance. We compared the approach with a more traditional data augmentation method of duplicating the training set speakers by adding noise \cite{hannun2014deep}\cite{tu2017objective}. Here we added white noise (SNR = 10dB) to the training samples. Figure \ref{fig:class_acc} shows the classification accuracies when adding a different number of augmented speakers to the original dataset. We  see that the performance gradually improves as additional speakers are added, and the proposed data augmentation method significantly outperformed the augmentation strategy of simply duplicating the training speakers and adding noise.

\section{Conclusion}
\label{sec:discussion}
In this paper, we proposed a new data augmentation strategy for clinical speech applications by transforming healthy speech to dysarthric speech. Our objective and subjective evaluation shows that the generated speech are acoustically and perceptually similar to real dysarthric speech, and the pilot classification experiment provides evidence that the augmentation strategy helps improve performance. However, further study is required to determine the benefits of the simulated speech samples in building large scale machine learning models. Future work will focus on reducing the perceptual artifacts by exploring other deep learning models that have been shown to be effective in voice transformation (e.g. recurrent neural networks (RNN)) \emph{and} using the resulting data to build larger machine learning systems.

\vfill\pagebreak



\footnotesize
\begin{thebibliography}{10}

\bibitem{yu2017recent}
Dong Yu and Jinyu Li,
\newblock ``Recent progresses in deep learning based acoustic models,''
\newblock {\em IEEE/CAA Journal of Automatica Sinica}, vol. 4, no. 3, pp.
  396--409, 2017.

\bibitem{tu2016relationship}
Ming Tu, Alan Wisler, Visar Berisha, and Julie~M Liss,
\newblock ``The relationship between perceptual disturbances in dysarthric
  speech and automatic speech recognition performance,''
\newblock {\em The Journal of the Acoustical Society of America}, vol. 140, no.
  5, pp. EL416--EL422, 2016.

\bibitem{fonseca2007wavelet}
Everthon~Silva Fonseca, Rodrigo~Capobianco Guido, Paulo~Rog{\'e}rio Scalassara,
  Carlos~Dias Maciel, and Jos{\'e}~Carlos Pereira,
\newblock ``Wavelet time-frequency analysis and least squares support vector
  machines for the identification of voice disorders,''
\newblock {\em Computers in Biology and Medicine}, vol. 37, no. 4, pp.
  571--578, 2007.

\bibitem{schuller2012interspeech}
Bj{\"o}rn~W Schuller, Stefan Steidl, Anton Batliner, Elmar N{\"o}th, Alessandro
  Vinciarelli, Felix Burkhardt, Rob Van~Son, Felix Weninger, Florian Eyben,
  Tobias Bocklet, et~al.,
\newblock ``The interspeech 2012 speaker trait challenge.,''
\newblock in {\em Interspeech}, 2012, vol. 2012, pp. 254--257.

\bibitem{kim2015automatic}
Jangwon Kim, Naveen Kumar, Andreas Tsiartas, Ming Li, and Shrikanth~S
  Narayanan,
\newblock ``Automatic intelligibility classification of sentence-level
  pathological speech,''
\newblock {\em Computer speech \& language}, vol. 29, no. 1, pp. 132--144,
  2015.

\bibitem{middag2009automated}
Catherine Middag, Jean-Pierre Martens, Gwen Van~Nuffelen, and Marc De~Bodt,
\newblock ``Automated intelligibility assessment of pathological speech using
  phonological features,''
\newblock {\em EURASIP Journal on Advances in Signal Processing}, vol. 2009,
  pp. 3, 2009.

\bibitem{rosen2000automatic}
Kristin Rosen and Sasha Yampolsky,
\newblock ``Automatic speech recognition and a review of its functioning with
  dysarthric speech,''
\newblock {\em Augmentative and Alternative Communication}, vol. 16, no. 1, pp.
  48--60, 2000.

\bibitem{green2003automatic}
Phil~D Green, James Carmichael, Athanassios Hatzis, Pam Enderby, Mark~S Hawley,
  and Mark Parker,
\newblock ``Automatic speech recognition with sparse training data for
  dysarthric speakers.,''
\newblock in {\em INTERSPEECH}, 2003.

\bibitem{jiao2015convex}
Yishan Jiao, Visar Berisha, Ming Tu, and Julie Liss,
\newblock ``Convex weighting criteria for speaking rate estimation,''
\newblock {\em IEEE/ACM transactions on audio, speech, and language
  processing}, vol. 23, no. 9, pp. 1421--1430, 2015.

\bibitem{jiao2017interpretable}
Yishan Jiao, Visar Berisha, and Julie Liss,
\newblock ``Interpretable phonological features for clinical applications,''
\newblock in {\em Acoustics, Speech and Signal Processing (ICASSP), 2017 IEEE
  International Conference on}. IEEE, 2017, pp. 5045--5049.

\bibitem{menendez1996nemours}
Xavier Menendez-Pidal, James~B Polikoff, Shirley~M Peters, Jennie~E Leonzio,
  and H~Timothy Bunnell,
\newblock ``The {Nemours} database of dysarthric speech,''
\newblock in {\em Spoken Language, 1996. ICSLP 96. Proceedings., Fourth
  International Conference on}. IEEE, 1996, vol.~3, pp. 1962--1965.

\bibitem{rudzicz2012torgo}
Frank Rudzicz, Aravind~Kumar Namasivayam, and Talya Wolff,
\newblock ``The {TORGO} database of acoustic and articulatory speech from
  speakers with dysarthria,''
\newblock {\em Language Resources and Evaluation}, vol. 46, no. 4, pp.
  523--541, 2012.

\bibitem{berisha2017float}
Visar Berisha, Julie Liss, Timothy Huston, Alan Wisler, Yishan Jiao, and
  Jonathan Eig,
\newblock ``Float like a butterfly sting like a bee: Changes in speech preceded
  {Parkinsonism} diagnosis for {Muhammad} {Ali},''
\newblock {\em Proc. Interspeech 2017}, pp. 1809--1813, 2017.

\bibitem{rudzicz2011articulatory}
Frank Rudzicz,
\newblock ``Articulatory knowledge in the recognition of dysarthric speech,''
\newblock {\em IEEE Transactions on Audio, Speech, and Language Processing},
  vol. 19, no. 4, pp. 947--960, 2011.

\bibitem{tu2017objective}
Ming Tu, Visar Berisha, and Julie Liss,
\newblock ``Objective assessment of pathological speech using distribution
  regression,''
\newblock in {\em Acoustics, Speech and Signal Processing (ICASSP), 2017 IEEE
  International Conference on}. IEEE, 2017, pp. 5050--5054.

\bibitem{tu2017interpretable}
Ming Tu, Visar Berisha, and Julie Liss,
\newblock ``Interpretable objective assessment of dysarthric speech based on
  deep neural networks,''
\newblock {\em Proc. Interspeech 2017}, pp. 1849--1853, 2017.

\bibitem{kain1998spectral}
Alexander Kain and Michael~W Macon,
\newblock ``Spectral voice conversion for text-to-speech synthesis,''
\newblock in {\em Acoustics, Speech and Signal Processing, 1998. Proceedings of
  the 1998 IEEE International Conference on}. IEEE, 1998, vol.~1, pp. 285--288.

\bibitem{kain2007improving}
Alexander~B Kain, John-Paul Hosom, Xiaochuan Niu, Jan~PH van Santen, Melanie
  Fried-Oken, and Janice Staehely,
\newblock ``Improving the intelligibility of dysarthric speech,''
\newblock {\em Speech communication}, vol. 49, no. 9, pp. 743--759, 2007.

\bibitem{rudzicz2011acoustic}
Frank Rudzicz,
\newblock ``Acoustic transformations to improve the intelligibility of
  dysarthric speech,''
\newblock in {\em Proceedings of the Second Workshop on Speech and Language
  Processing for Assistive Technologies}. Association for Computational
  Linguistics, 2011, pp. 11--21.

\bibitem{hsu2017voice}
Chin-Cheng Hsu, Hsin-Te Hwang, Yi-Chiao Wu, Yu~Tsao, and Hsin-Min Wang,
\newblock ``Voice conversion from unaligned corpora using variational
  autoencoding wasserstein generative adversarial networks,''
\newblock {\em arXiv preprint arXiv:1704.00849}, 2017.

\bibitem{kaneko2017generative_1}
Takuhiro Kaneko, Hirokazu Kameoka, Nobukatsu Hojo, Yusuke Ijima, Kaoru
  Hiramatsu, and Kunio Kashino,
\newblock ``Generative adversarial network-based postfilter for statistical
  parametric speech synthesis,''
\newblock in {\em Proc. 2017 IEEE International Conference on Acoustics, Speech
  and Signal Processing (ICASSP2017)}, 2017, pp. 4910--4914.

\bibitem{kaneko2017generative_2}
Takuhiro Kaneko, Shinji Takaki, Hirokazu Kameoka, and Junichi Yamagishi,
\newblock ``Generative adversarial network-based postfilter for {STFT}
  spectrograms,''
\newblock in {\em Proceedings of Interspeech}, 2017.

\bibitem{duffy2013motor}
Joseph~R Duffy,
\newblock {\em Motor Speech Disorders: Substrates, Differential Diagnosis, and
  Management},
\newblock Elsevier Health Sciences, 2013.

\bibitem{toolkit}
Ramon Corretge,
\newblock ``Praat vocal toolkit: A praat plugin with automated scripts for
  voice processing,'' http://www.praatvocaltoolkit.com/, 2012,
\newblock [Computer software].

\bibitem{kawahara1999restructuring}
Hideki Kawahara, Ikuyo Masuda-Katsuse, and Alain De~Cheveigne,
\newblock ``Restructuring speech representations using a pitch-adaptive
  time--frequency smoothing and an instantaneous-frequency-based f0 extraction:
  Possible role of a repetitive structure in sounds,''
\newblock {\em Speech communication}, vol. 27, no. 3, pp. 187--207, 1999.

\bibitem{sptk2009speech}
SPTK~Working Group et~al.,
\newblock ``Speech signal processing toolkit (sptk),''
\newblock {\em http://sp-tk. sourceforge. net}, 2009.

\bibitem{goodfellow2014generative}
Ian Goodfellow, Jean Pouget-Abadie, Mehdi Mirza, Bing Xu, David Warde-Farley,
  Sherjil Ozair, Aaron Courville, and Yoshua Bengio,
\newblock ``Generative adversarial nets,''
\newblock in {\em Advances in neural information processing systems}, 2014, pp.
  2672--2680.

\bibitem{radford2015unsupervised}
Alec Radford, Luke Metz, and Soumith Chintala,
\newblock ``Unsupervised representation learning with deep convolutional
  generative adversarial networks,''
\newblock {\em arXiv preprint arXiv:1511.06434}, 2015.

\bibitem{abadi2016tensorflow}
Mart{\'\i}n Abadi, Ashish Agarwal, Paul Barham, Eugene Brevdo, Zhifeng Chen,
  Craig Citro, Greg~S Corrado, Andy Davis, Jeffrey Dean, Matthieu Devin,
  et~al.,
\newblock ``Tensorflow: Large-scale machine learning on heterogeneous
  distributed systems,''
\newblock {\em arXiv preprint arXiv:1603.04467}, 2016.

\bibitem{kingma2014adam}
Diederik Kingma and Jimmy Ba,
\newblock ``Adam: A method for stochastic optimization,''
\newblock {\em arXiv preprint arXiv:1412.6980}, 2014.

\bibitem{liss2009quantifying}
Julie~M Liss, Laurence White, Sven~L Mattys, Kaitlin Lansford, Andrew~J Lotto,
  Stephanie~M Spitzer, and John~N Caviness,
\newblock ``Quantifying speech rhythm abnormalities in the dysarthrias,''
\newblock {\em Journal of Speech, Language, and Hearing Research}, vol. 52, no.
  5, pp. 1334--1352, 2009.

\bibitem{berisha2016empirically}
Visar Berisha, Alan Wisler, Alfred~O Hero, and Andreas Spanias,
\newblock ``Empirically estimable classification bounds based on a
  nonparametric divergence measure,''
\newblock {\em IEEE Transactions on Signal Processing}, vol. 64, no. 3, pp.
  580--591, 2016.

\bibitem{wisler2014domain}
Alan Wisler, Visar Berisha, Julie Liss, and Andreas Spanias,
\newblock ``Domain invariant speech features using a new divergence measure,''
\newblock in {\em Spoken Language Technology Workshop (SLT), 2014 IEEE}. IEEE,
  2014, pp. 77--82.

\bibitem{hannun2014deep}
Awni Hannun, Carl Case, Jared Casper, Bryan Catanzaro, Greg Diamos, Erich
  Elsen, Ryan Prenger, Sanjeev Satheesh, Shubho Sengupta, Adam Coates, et~al.,
\newblock ``Deep speech: Scaling up end-to-end speech recognition,''
\newblock {\em arXiv preprint arXiv:1412.5567}, 2014.

\end{thebibliography}
\end{document}